\documentclass[iop]{emulateapj}
\usepackage{amssymb,amsmath,epsfig,times,longtable,color,textcomp}

\newcommand{\rxte}{\textit{RXTE}}

\newcommand{\chandra}{\textit{Chandra}}
\newcommand{\nustar}{\textit{NuSTAR}}
\newcommand{\swift}{\textit{Swift}}
\newcommand{\suzaku}{{\it Suzaku}}
\newcommand{\xmm}{{\it XMM-Newton}}

\begin{document}

\lefthead{\nustar\ observations of the very high state in GX~339-4}
\righthead{Parker et al.}

\title[NuSTAR observations of the very-high state in GX~339-4]{NuSTAR and Swift observations of the very high state in GX~339-4: weighing the black hole with X-rays}

\author{
M. L. Parker\altaffilmark{1},
J. A. Tomsick\altaffilmark{2},
J. A. Kennea\altaffilmark{3},
J. M. Miller\altaffilmark{4},
F. A. Harrison\altaffilmark{5},
D. Barret\altaffilmark{6},
S. E. Boggs\altaffilmark{2},
F. E. Christensen\altaffilmark{7},
W. W. Craig\altaffilmark{2,8},
A. C. Fabian\altaffilmark{1},
F. F\"urst\altaffilmark{5}
V. Grinberg\altaffilmark{9},
C. J. Hailey\altaffilmark{10},
P. Romano\altaffilmark{11}
D. Stern\altaffilmark{12},
D. J. Walton\altaffilmark{12}
and W. W. Zhang\altaffilmark{13}
}

                                              
\altaffiltext{1}{Institute of Astronomy, Madingley Road, Cambridge, CB3 0HA, UK}                      
\altaffiltext{2}{Space Sciences Laboratory, University of California, Berkeley, 7 Gauss Way, Berkeley, CA 94720-7450, USA}  
\altaffiltext{3}{Department of Astronomy and Astrophysics, Pennsylvania State University, 525 Davey Lab, University Park, PA 16802, USA}
\altaffiltext{4}{Department of Astronomy, University of Michigan, 1085 South University Avenue, West Hall 311, Ann Arbor, MI 48109-1042, USA}
\altaffiltext{5}{California Institute of Technology, 1200 East California Boulevard, Pasadena, CA 91125, USA}
\altaffiltext{6}{Institut de Recherche en Astrophysique et Plan\'etologie, 9 Avenue du Colonel Roche, 31028 Toulouse, France}
\altaffiltext{7}{Danish Technical University, DK-2800 Lyngby, Denmark}
\altaffiltext{8}{Lawrence Livermore National Laboratory, Livermore, CA, USA}
\altaffiltext{9}{Massachusetts Institute of Technology, Kavli Institute for Astrophysics, Cambridge, MA 02139, USA}
\altaffiltext{10}{Columbia University, New York, NY 10027, USA}
\altaffiltext{11}{INAF-IASF Palermo, Via Ugo La Malfa 153, 90146 Palermo, Italy}
\altaffiltext{12}{Jet Propulsion Laboratory, California Institute of Technology, 4800 Oak Grove Drive, Pasadena, CA 91109, USA}
\altaffiltext{13}{NASA Goddard Space Flight Center, Greenbelt, MD 20771, USA}       

\begin{abstract}
We present results from spectral fitting of the very high state of GX~339-4 with \nustar\ and \swift . We use relativistic reflection modeling to measure the spin of the black hole and inclination of the inner disk, and find a spin of $a=0.95^{+0.02}_{-0.08}$ and inclination of $30$\textdegree$\pm1$ (statistical errors). These values agree well with previous results from reflection modelling. 
With the exceptional sensitivity of \nustar\ at the high-energy side of the disk spectrum, we are able to constrain multiple physical parameters simultaneously using continuum fitting. By using the constraints from reflection as input for the continuum fitting method, we invert the conventional fitting procedure to estimate the mass and distance of GX~339-4 using just the X-ray spectrum, finding a mass of $9.0^{+1.6}_{-1.2}M_\odot$ and distance of $8.4\pm0.9$~kpc (statistical errors).
\end{abstract}

\keywords{
accretion, accretion disks -- X-rays: binaries -- X-rays: individual: GX~339-4 -- black hole physics
}

\section{Introduction}

There are two methods of determining the dimensionless black hole spin parameter, $a$, in X-ray binaries. The first relies on measuring the relativistic distortion of the iron K$\alpha$ line, originating in reflection from the inner accretion disk \citep{Fabian89}. The second uses the spectrum of the disk itself, which is strongly dependent on the inner radius \citep{Zhang97}. Both methods have been used successfully in several X-ray binaries \citep[see reviews by ][]{McClintock14, Miller_and_Miller15, Middleton15}, however the results are not always consistent between the two \citep{Kolehmainen14}. Some of this discrepancy can be attributed to the difficulty of using both methods simultaneously, because they rely on being able to measure different spectral components. The optimal state for measuring the iron line may not be optimal for measuring the disk spectrum, and vice versa. A second difficulty arises from pile-up, which can potentially introduce large uncertainties, the extent of which are difficult to establish, and can lead to conflicting results \citep[e.g. ][]{Ng10,Miller10}.

The launch of the \emph{Nuclear Spectroscopic Telescope Array} \citep[\nustar ;][]{Harrison13} has revolutionized the study of reflection in X-ray binaries. \nustar\ has a triggered read-out, meaning that it does not suffer from pile-up when looking at bright sources (pile-up occurs when two or more photons arrive close together and are read as a single event). \nustar\ thus allows for reliable, sensitive, low background spectroscopy from 3--79~keV, covering both the iron line and Compton hump from X-ray reflection. This has enabled precision measurements of the reflection parameters in several X-ray binaries \citep{Miller13,Miller13Serpens,Tomsick14,King14,Fuerst15,Parker15_cygx1}.

GX~339-4 is a well known transient X-ray binary. It was discovered in 1973 by \citet{Markert73}, who noted that it changed by a factor of $\gtrsim60$ in flux over a year. It has since been studied extensively, as it is a very bright source and frequently undergoes outbursts. There have been several attempts to measure the spin in GX~339-4. \citet{Miller04_339XMM,Miller04_339Chandra} used \xmm\ and \rxte , and the \chandra\ HETGS, to measure a relativistically broadened iron line. From the \xmm /\rxte\ spectrum \citet{Miller04_339XMM} argue for a high spin of $a>0.8$--0.9. \citet{Reis08_339} re-analysed the archival \xmm\ and \rxte\ data using new reflection models, and find a spin of $a=0.935\pm0.01$. \citet{Miller08_339} combine measurements with \suzaku\ and \xmm\ to find a best estimate of $a=0.94\pm0.01$ (statistical) $\pm0.04$ (systematic), in good agreement with the measurement of \citet{Reis08_339}. \citet{Yamada09_339} use the same \suzaku\ data as \citet{Miller08_339}, excising more of the core of the PSF to further limit the effects of pile-up, and find that the line appears narrower when more data are excluded. From this, they infer that the broadening of the line is due to pile-up, and that GX~339-4 need not be rapidly spinning (although their measurements are still consistent with maximal spin at the 90\% confidence level). \citet{Kolehmainen10} use continuum fitting and physical arguments based on the inclination and distance of the binary system to argue that the spin must be less than 0.9. Most recently, \citet{Ludlam15_339} re-analyse the data from \citet{Miller04_339XMM} and \citet{Miller08_339} and find $a>0.97$, and \citet{Garcia15_gx339} find $a=0.93^{+0.03}_{-0.05}$ from stacked RXTE data.

The inclination of the GX~339-4 binary system is only weakly constrained by dynamic measurements. As the system  is non-eclipsing, the inclination must be less than 60\textdegree\ \citep{Cowley02}, and a plausible lower limit of $i\gtrsim 45$\textdegree\ is given from the mass function \citep{Zdziarski04}. 
However, measurements of the inner disk inclination from the broad iron line generally give lower values. \citet{Miller04_339XMM} find $i=12$\textdegree$_{-2}^{+4}$; \citet{Miller08_339} find $i=19$\textdegree$\pm1$; and \citet{Reis08_339} find $i=18.2$\textdegree$_{-0.5}^{+0.3}$. \citet{Cassatella12} use a combined spectroscopy and timing analysis to find $i<30$\textdegree, fitting both the lag and spectral data. More recently, \citet{Fuerst15} used five hard state observations from the failed outburst in 2013 to constrain the reflection spectrum, finding inclinations ranging from 31\textdegree\ to 59\textdegree , depending on the model. \citet{Ludlam15_339} find an inclination of $36$\textdegree$\pm4$\textdegree\ using the latest \textsc{relxill} reflection models \citep{Garcia14}, which self-consistently take into account the inclination angle. \citet{Garcia15_gx339} also use \textsc{relxill}, combined with extremely high signal but low resolution stacked \rxte\ spectra, to find an inclination of $48$\textdegree$\pm1$.

In this paper we use broad-band spectroscopy with \swift\ and \nustar\ to constrain the spin and inclination using relativistic reflection. We then use these estimates as input for the continuum fitting model, and fit for the mass and distance, obtaining a constraint on these parameters using just the X-ray spectrum.

\section{Observations and Data Reduction}
\label{section_datareduction}

\swift\ monitoring detected strong thermal and power-law components in the spectrum of GX~339-4 on 2015 March 4th, suggesting that both reflection and continuum fitting methods could be used simultaneously. This state is frequently referred to as the very high state \citep[or steep power-law state; see review by][]{McClintock06}. A \nustar\ target of opportunity (TOO) was triggered, with a simultaneous \swift\ snapshot, for clean on-source exposures of $\sim2$ and $30$~ks, respectively. 

The {\it Swift}/XRT data were processed with standard procedures
(\texttt{xrtpipeline} v0.13.1),
filtering and screening criteria using \texttt{FTOOLS} (v6.16).
The data, collected in windowed-timing mode, were affected by pileup. Following \citet{Romano06},
source events were accumulated within an annular region with an
outer radius of 20 pixels (1 pixel $\sim2\farcs36$) and inner radius
of 10 pixels.
Background events were accumulated from a source-free region nearby.
For our spectral analysis, ancillary response files were generated
with \texttt{xrtmkarf}.
We used the XRT spectral redistribution matrices in CALDB (20140709). We bin the spectrum to a signal-to-noise ratio of 30, after background subtraction, and fit from 1--4~keV.

The \nustar\ data (ObsID 80001015003) were reduced using \nustar\ Data Analysis Software (NuSTARDAS) 1.4.1 and CALDB version 20150316. Source spectra were extracted from 150$^{\prime\prime}$ circular extraction regions centered on the source position, and background spectra were extracted from $\sim100^{\prime\prime}$ circular regions from the opposite corner of the detector (the least contaminated with source photons). The source count rate is a factor of 10 or more above the background at all energies. We bin the \nustar\ FPMA and FPMB spectra to oversample the spectrum by a factor of 3, and to a signal-to-noise ratio of 50. We fit the spectra over the whole energy range (3--79~keV); however, the final spectral bin is extremely large due to the steep spectrum and extends past 79~keV, so we exclude it. This gives an effective upper limit of $\sim60$~keV. 

All errors are 1$\sigma$ unless otherwise stated. All spectral fitting is done in \textsc{Xspec} 12.9.0. In all cases we use \emph{wilm} abundances \citep{Wilms00}, and \emph{vern} cross-sections \citep{Verner96}.

\section{Results}
\label{section_results}

\begin{figure}
\centering
\includegraphics[width=\linewidth]{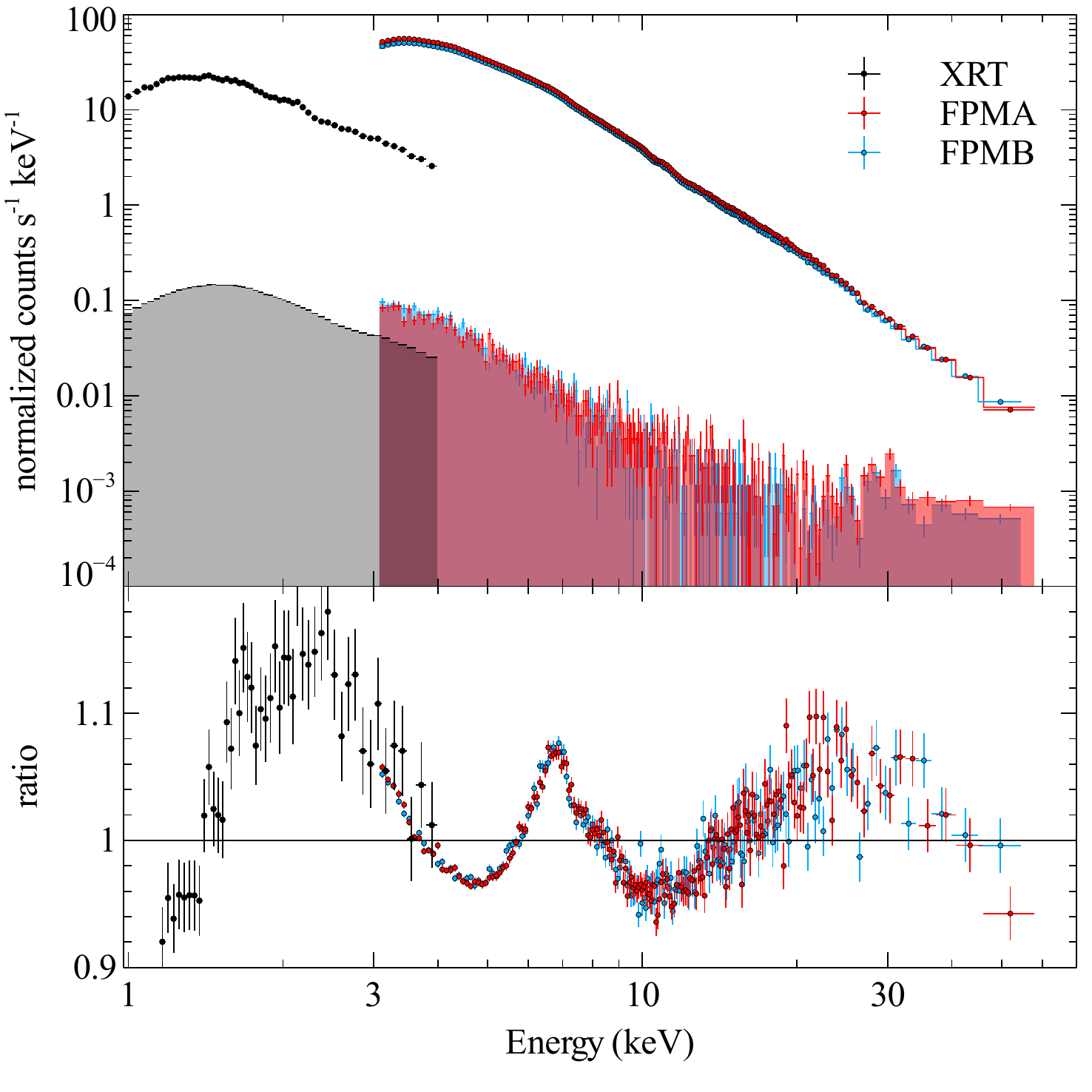}
\caption{Top: X-ray spectrum of GX~339-4, fit with an absorbed power-law and disk black body. Shaded regions show the background spectra for each instrument. Bottom: Residuals to the model.}
\label{fig_powerlaw_resids}
\end{figure}

In Fig.~\ref{fig_powerlaw_resids} we show the residuals to an absorbed power-law/disk black body spectrum (\emph{tbabs*[diskbb+powerlaw]} in \textsc{xspec}). The broad iron line and Compton hump are clearly visible, and there is a prominent excess below $\sim3$~keV where the simple phenomenological model does not adequately describe the data.

We fit the combined \nustar /XRT spectrum with a three-component disk plus Comptonization plus reflection model. We use \textit{kerrbb} \citep{Li05_kerrbb} for the disk spectrum, \emph{comptt} \citep{Titarchuk94} for the Comptonization, and \emph{relxilllp} \citep{Garcia14} for the relativistic reflection. As discussed in \citet{Parker15_cygx1}, the \emph{relxilllp} lamp-post model has the advantage of parameterising the emissivity profile in physical units, requiring a smaller number of parameters, and restricting the profile to physically plausible regions of parameter space. In addition, we allow the photon index of the reflection spectrum to vary independently of the continuum, as required in \citet{Fuerst15}. We tie the high energy cutoff of the reflection spectrum to twice the plasma temperature of \emph{comptt} \citep{Petrucci01}. We tie the spin and inclination parameters of the disk and reflection components together, but leave the mass and distance parameters of \emph{kerrbb} free to vary, as these are largely unconstrained \citep[see e.g.][]{Hynes03}. This differs from the conventional continuum fitting procedure, where the mass, distance, and inclination parameters are fixed and used to constrain the spin.
The \emph{relxilllp} model has a parameter for the reflection fraction, which allows for a powerlaw continuum. As we already include a Comptonization continuum we fix this parameter to -1, i.e. no continuum emission. 
We leave the color-correction factor at the default value of 1.7. 
Finally, we allow for a constant offset between the three instruments, to account for differences in flux calibration, and allow for a slight difference in $\Gamma$ between FPMA and FPMB \citep[this is a $<$1\% effect, within the known calibration uncertainties;][]{Madsen15}.

\begin{figure}
\centering
\includegraphics[width=\linewidth]{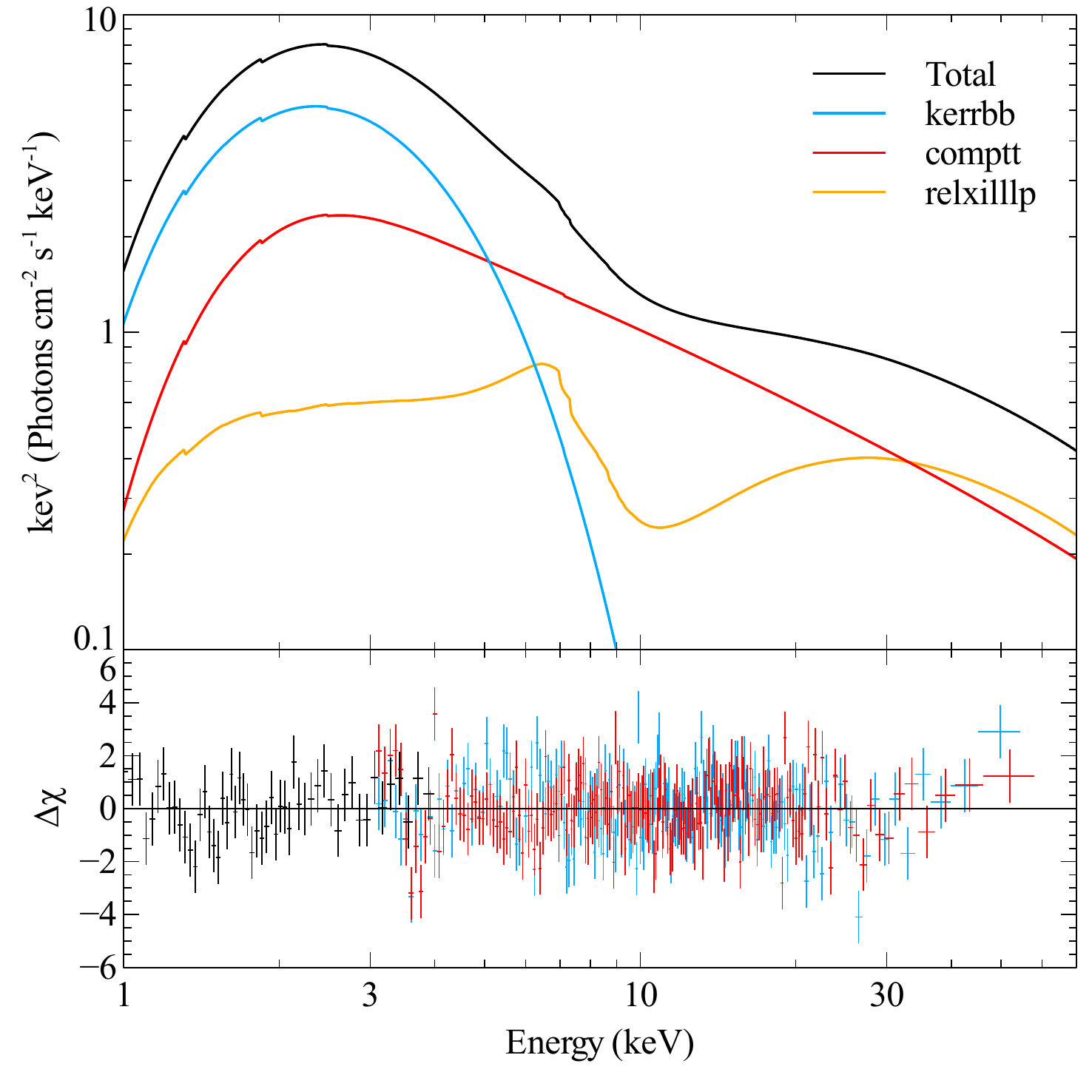}
\caption{Top: Best fit model, showing the different model components. Bottom: Best fit residuals. Note that we plot the residuals in units of $\Delta\chi$, so that the XRT and \nustar\ residuals are comparable in magnitude.}
\label{fig_bestfit}
\end{figure}

This model gives a reasonable fit to the data ($\chi^2_\nu=1.32$), given the multiple instruments/detectors and extremely high count rate, and no significant residuals are visible. The best fit parameters are given in Table~\ref{table_fitpars}, and the model and residuals are shown in Fig.~\ref{fig_bestfit}. The mass and distance parameters are well constrained, as are the spin and inclination.

\begin{figure*}
\centering
\includegraphics[width=10cm]{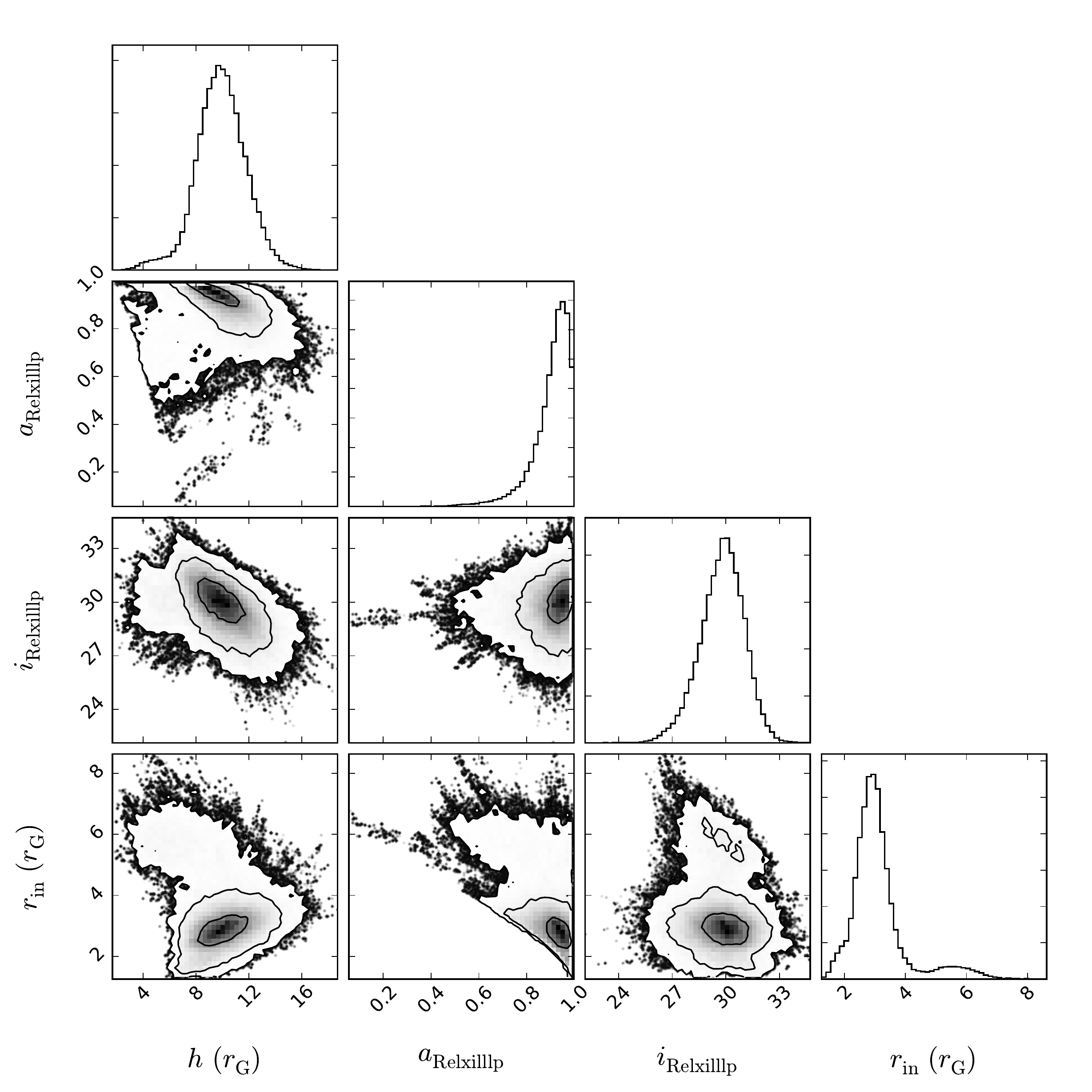}
\includegraphics[width=5.65cm]{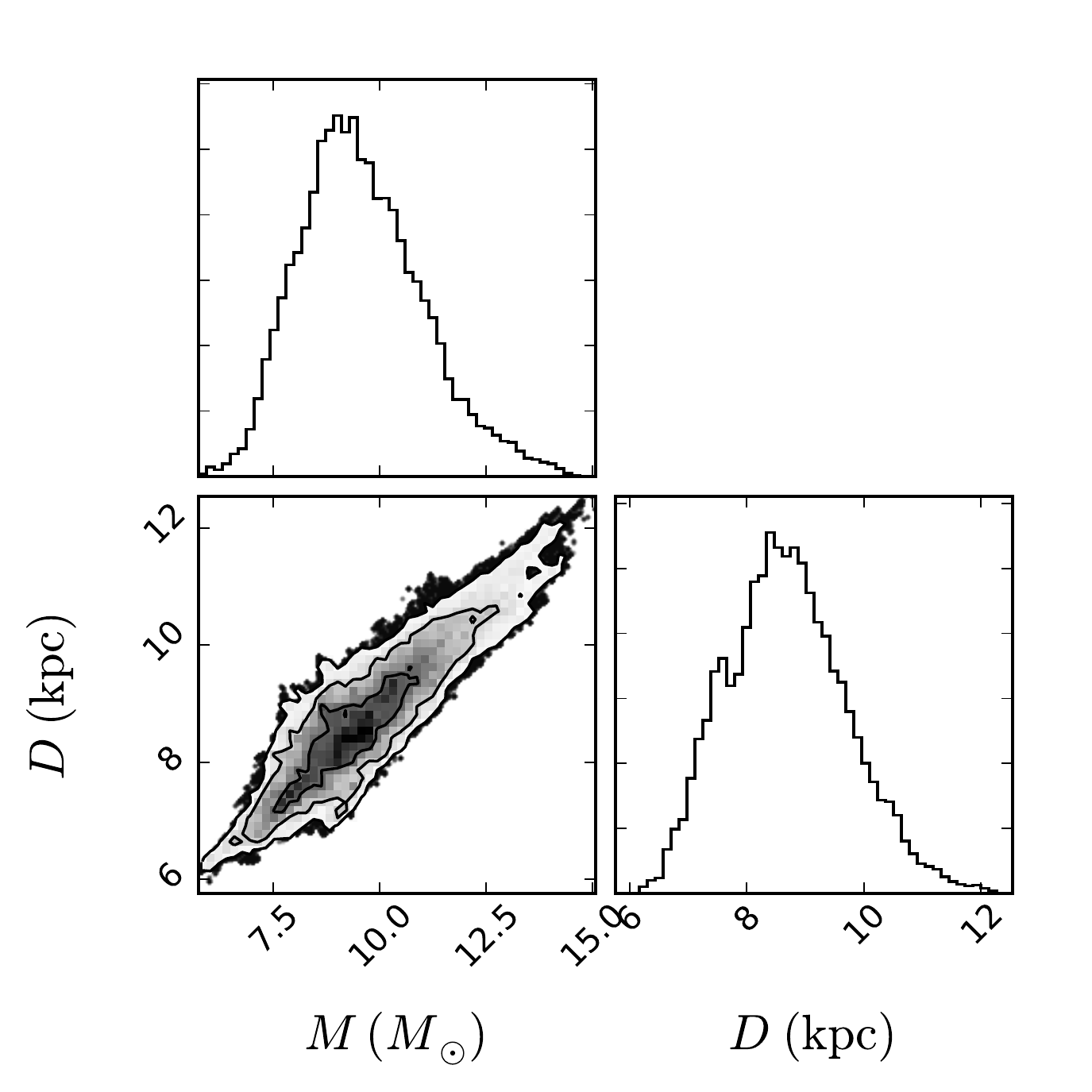}
\caption{Left: 1,2 and 3$\sigma$ confidence contours for spin, inclination, source height, and inner disk radius measured using reflection spectroscopy. Source height and inner disk radius are required to be $\geqslant r_\textrm{ISCO}$ Right: contours for the mass and distance measured using continuum fitting.}
\label{fig_contours}
\end{figure*}

As in \citet{Parker15_cygx1}, we use the \textsc{xspec\_emcee} (written by Jeremy Sanders) implementation of the \textsc{emcee} MCMC code \citep{Foreman-Mackey13} to calculate confidence contours for the model parameters and search for degeneracies. We use 500 walkers for 10000 steps each, with an initial burn-in period of 1000 steps\footnote{The autocorellation time is $\sim100$ steps. The 1000 step burn-in time thus covers $\sim10$ autocorellation times, sufficient given the large number of walkers. The effective sample size is then $\sim50000$.}.  We show these contours for selected parameters of interest in Fig.~\ref{fig_contours}.

\begin{table}
\begin{center}

\caption{Best fit model parameters.}
\label{table_fitpars}
\begin{tabular}{l c l}
\hline
\hline
Parameter & Value & Unit/Description\\
\hline
$N_\textrm{H}$ & $(7.7\pm0.2)\times10^{21}$ & Column density (cm$^{-2}$)\\
$M$ & $9.0^{+1.6}_{-1.2}$ & Black hole mass ($M_\odot$)\\
$D$ & $8.4\pm0.9$ & Distance (kpc)\\
$a$ & $0.95^{+0.02}_{-0.08}$ & Spin\\
$i$ & $30\pm1$ & Inclination (degrees)\\
$\dot{M}$ & $(7.6\pm0.4)\times10^{17}$ & Accretion rate (g~s$^{-1}$)\\
$T_0$ & $0.38\pm0.01$ & Seed temperature (keV)\\
$T_\textrm{plasma}$ & $183^{+3}_{-11}$ & Plasma temperature (keV)\\
$\tau$ & $<1.07\times10^{-2}$ & Optical depth\\
$h$ & $9\pm2$ & Source height ($r_\textrm{G}$)\\
$r_\textrm{in}$ & $<1.5$ & Inner radius ($r_\textrm{ISCO}$)\\
$\Gamma$ & $2.12\pm0.02$ & Reflection photon index\\
$A_\textrm{Fe}$ & $6.6_{-0.6}^{+0.5}$ & Iron abundance (solar)\\
$\log(\xi)$ & $3.65\pm0.05$ & Ionization parameter (erg~cm~s$^{-1}$)\\
$\chi^2$/dof & 538/407=1.32\\
\hline
$C_\textrm{FPMA}$ & 1 & Calibration constants\\
$C_\textrm{FPMB}$ & $1.0067\pm0.0007$\\
$C_\textrm{XRT}$ & $1.10\pm0.01$ & \\

\hline
\hline
\end{tabular}
\end{center}

Note. Errors are statistical only, and are calculated for each parameter from the MCMC after marginalising over all other parameters. \\
\end{table}

\begin{figure*}
\centering
\includegraphics[width=6cm,natwidth=8in,natheight=8in]{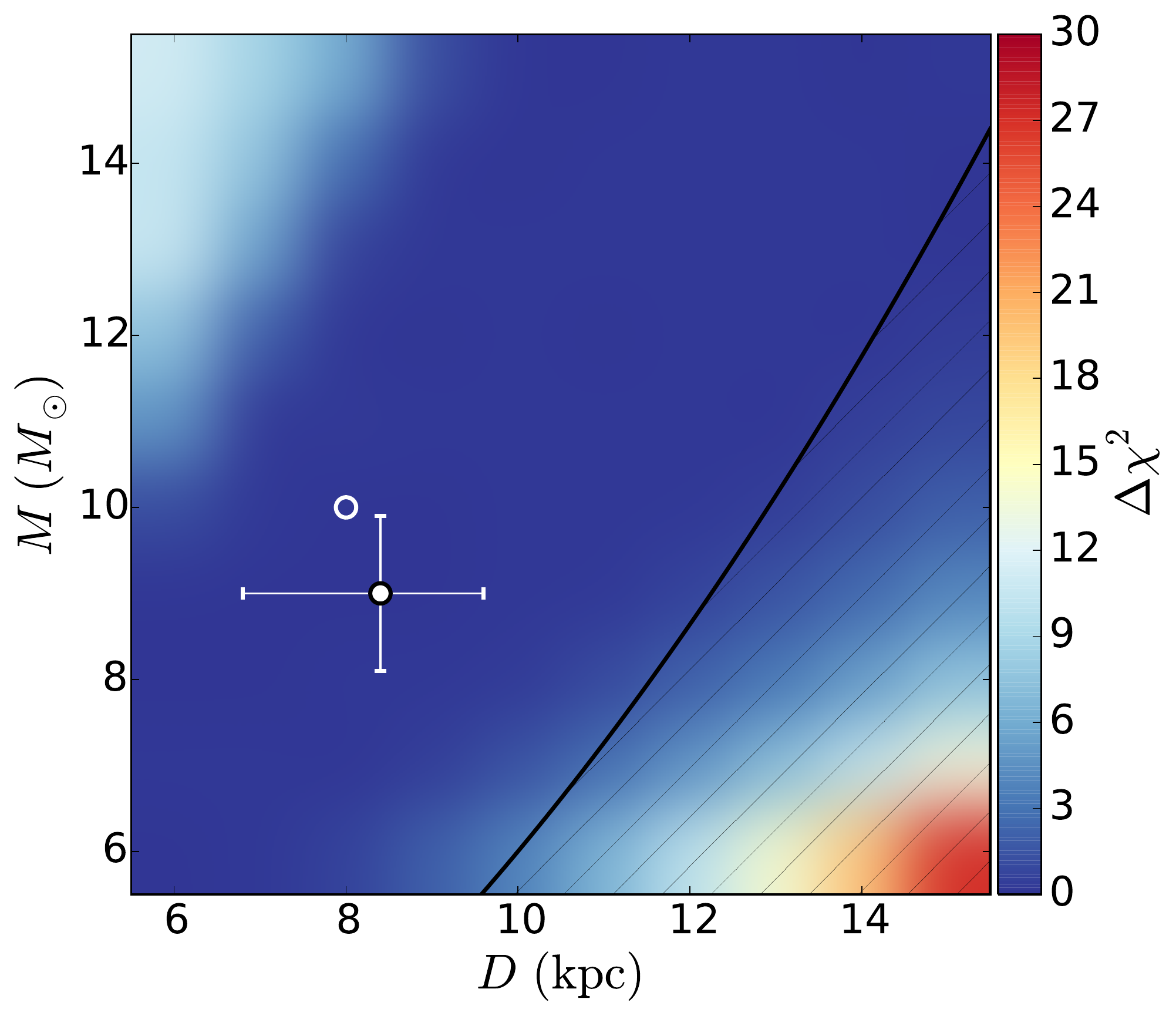}
\hspace{20pt}
\includegraphics[width=6cm,natwidth=8in,natheight=8in]{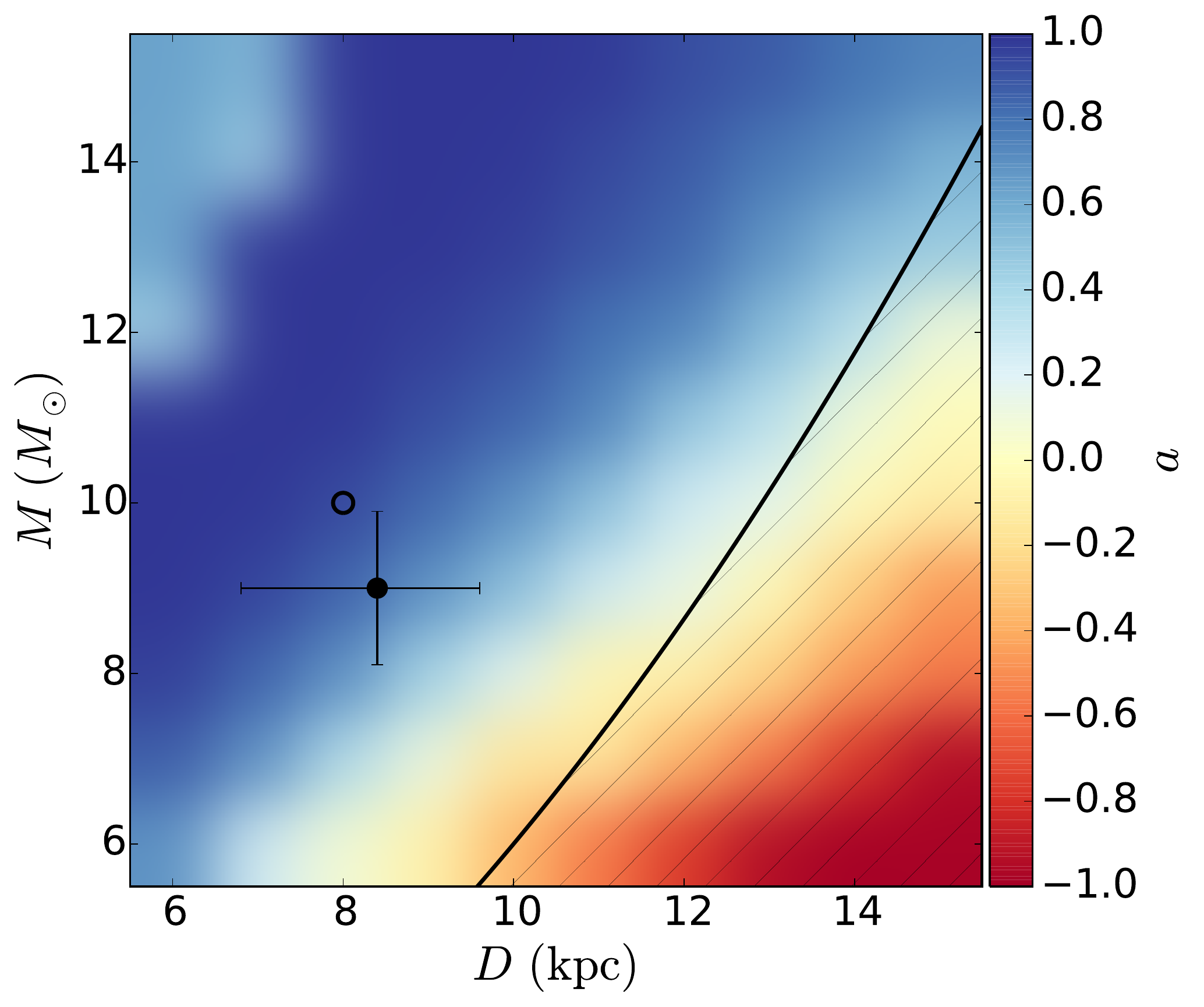}\\
\vspace{-20pt}
\includegraphics[width=6cm,natwidth=8in,natheight=8in]{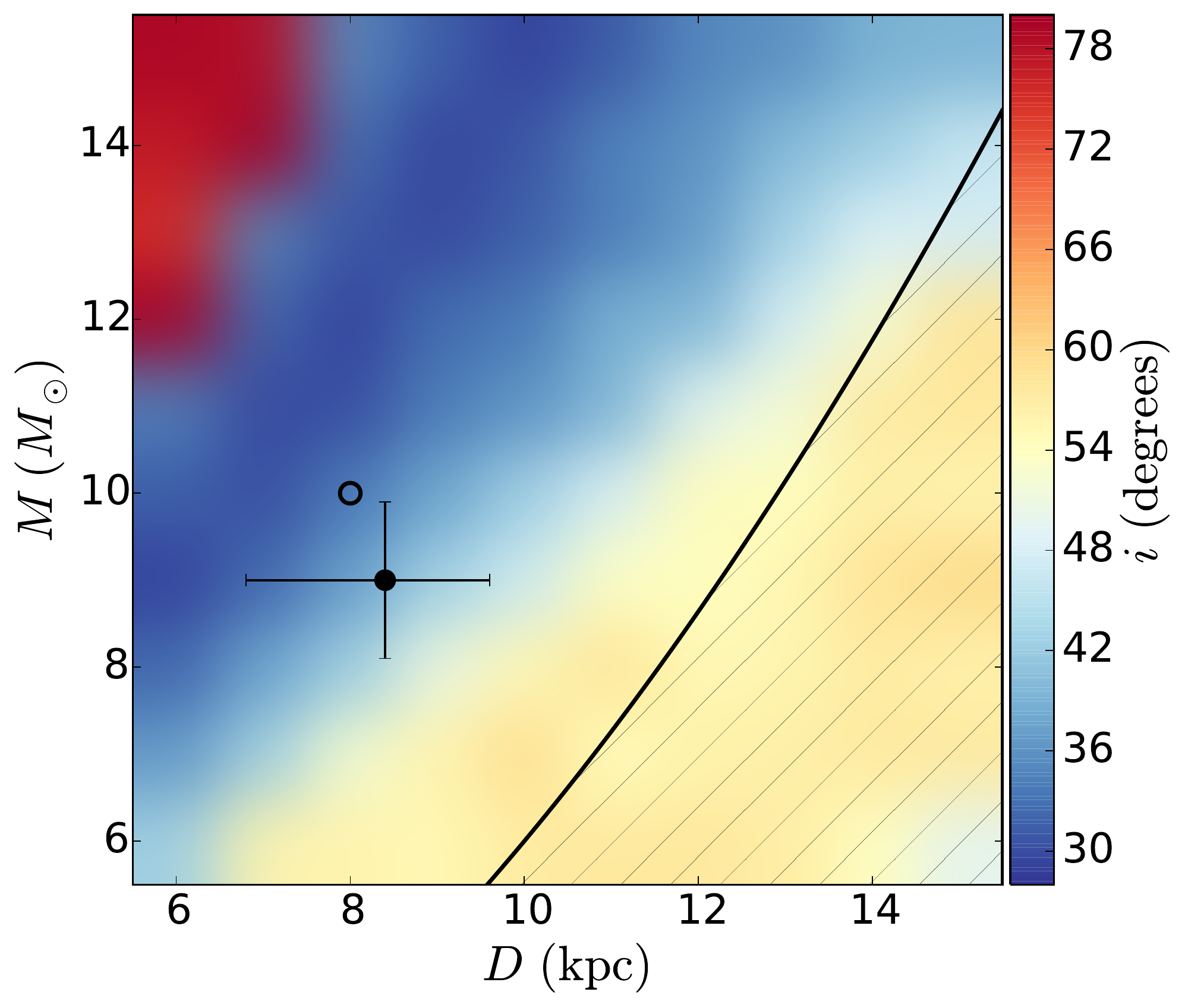}
\caption{Goodness-of-fit relative to best-fit model, and difference between the spin parameter and inclination from the continuum model compared to the reflection model, evaluated over the potential mass/distance parameter space and smoothed with a gaussian. The filled point shows our estimate of $M$ and $D$ from the X-ray spectrum, and the open point shows the value from \citet{Zdziarski04}. The hatched region shows where the bolometric luminosity exceeds $30\%$ of the Eddington luminosity and the continuum method is not valid.}
\label{fig_MandD}
\end{figure*}

We also perform a second fit, designed to establish the spin and inclination estimates from continuum fitting as a function of mass and distance. Distance estimates range from $\sim$6--15~kpc, and mass estimates cover the same range in solar masses. We therefore fit our model at every integer pair of $M$ and $D$ values within this range (the spin, inclination, and $\chi^2$ values are shown in Fig.~\ref{fig_MandD}). The fit remains good over most of the parameter space, and we find a band of solutions that are consistent with the reflection estimates of $i$ and $a$, with large deviations in the high $M$, low $D$ and low $M$, high $D$ corners.

\section{Discussion}
\label{section_discussion}

The data used in this work is some of the highest quality ever taken of GX~339-4, with higher spectral resolution than \rxte , and none of the issues arising from pile-up that affect \xmm\ and \chandra . We find a high spin value ($a=0.95^{+0.02}_{-0.08}$) and a low inclination ($30$\textdegree$\pm1$) from reflection fitting. These values are in good agreement with previous measurements of the spin from relativistic reflection, particularly those of \citet{Reis08_339},  \citet{Miller08_339}, and \citet{Garcia15_gx339}, all of whom found $a\sim0.94$. The consensus now firmly points towards a high spin ($a>0.9$), particularly with \nustar\ measurements where the observed broadening cannot be due to pile-up effects. The $a<0.9$ result of \citet{Kolehmainen10} relies on the inclination of the inner disk being the same as that of the binary system, which may not be the case.
Our measurement of the inclination is consistent with the majority of other recent results from reflection fitting \citep[e.g.][]{Plant14, Ludlam15_339}, although notably lower than that found by \citet{Garcia15_gx339}. This difference may be due to the lower spectral resolution of the data used, but more work is needed to determine the true inclination of the inner disk in GX~339-4.


By combining the reflection and continuum fitting methods, we have been able to constrain the mass and distance of GX~339-4 using just the X-ray spectrum. The mass and distance are strongly correlated (see Fig.~\ref{fig_MandD}), but not so degenerate that a meaningful constraint cannot be obtained. We are able to constrain multiple parameters (either $M$ and $D$ or $a$ and $i$) using continuum fitting because of the exceptional quality of the \nustar\ data covering the high energy side of the disk spectrum. The \emph{kerrbb} model is significantly contributing flux in the \nustar\ band up to 10~keV, and dominates the emission until $\sim5$~keV. The 3--10~keV band is both where \nustar\ is most sensitive and where the disk spectrum shows the greatest sensitivity to parameter changes, as it originates from the inner disk, so \nustar\ allows us to measure multiple parameters with precision.

There are various systematic effects that may impact both the reflection and continuum fitting methods. We have assumed a colour-correction factor of 1.7 for \emph{kerrbb}. We test two additional values, f=1.5 and 1.9 \citep[e.g.][]{Davis06}, and find masses of $7.8_{-0.1}^{+0.7}\ M_\odot$ and $11.3_{-0.2}^{+0.5}\ M_\odot$, respectively, and distances of $7.2_{-0.1}^{+0.2}$~kpc and $8.2^{+0.7}_{-0.2}$~kpc. This represents a systematic error of $\sim$1--2~$M_\odot$ in mass and $\sim1$~kpc in distance.
The continuum method is only valid when the source is accreting below $\sim30$\% of the Eddington limit. Due to the uncertainty in $M$ and $D$, this is hard to be certain of, but based on the X-ray luminosity we find that this condition is met when $M/M_\odot>0.06(D/1\mathrm{kpc})^2$. This condition is met for the majority of the parameter space concerned \citep[including the estimate of][and our estimate, see Fig.~\ref{fig_MandD}]{Zdziarski04}. Additionally, the continuum fitting method is only valid when the fraction of disk photons that are Compton scattered is $\lesssim0.25$ \citep{Steiner09}. In this case, the fraction is $\sim0.15$, and is not affected by mass or distance.

We rely heavily on measurements of the spectral shape of the disk at high energies with \nustar\ to constrain multiple parameters from continuum fitting. While the sensitivity of \nustar\ means that the disk is detectable up to $\sim10$~keV, the fraction of flux it contributes is low above 5~keV due to the steep spectral shape. This means that small differences in the dominant Comptonization component can potentially have a large impact on the measured parameters. To test this, we re-fit the data in the same way using the alternative Comptonization models \emph{nthcomp} \citep{Zycki99}, \emph{compps} \citep{Poutanen96}, and \emph{eqpair} \citep{Coppi99}. We find masses of $9.8\pm0.2\ M_\odot$, $11.2^{+0.7}_{-0.2}\ M_\odot$, and $8.8^{+0.5}_{-0.4}\ M_\odot$, respectively, and distances of $8.3^{+0.4}_{-0.3}$~kpc, $9.7_{-0.1}^{+0.9}$~kpc, and $8.8_{-0.2}^{+0.3}$~kpc\footnote{These errors and those for the different color-correction factors are calculated using the \textsc{Xspec} error algorithm, rather than MCMC, which is likely why they are smaller than those reported in Table~\ref{table_fitpars}.}. Taking the standard deviation of these measurements (including the original with \emph{comptt}) as an estimate of the systematic error gives $\pm1.1\ M_\odot$ in mass and $\pm0.6$~kpc in distance.

A final systematic issue affecting the continuum method is conserving the number of Compton scattered photons from the disk. We are unable to find an acceptable fit using the \emph{simpl} model \citep{Steiner09}, due to its lack of spectral curvature needed for the \nustar\ spectrum. In one case out of six considered by \citeauthor{Steiner09}, \emph{comptt} underpredicts the inner radius of the disk by 0.5~$R_\textrm{G}$ relative to \emph{simpl}. Translating this to spin (at the best fit value from reflection) gives a conservative systematic error of $0.95^{+0.04}_{-0.07}$, which is comparable to the statistical error on spin and therefore will not have a large effect on our results. In general, the choice of hard component model may have a systematic effect on the results from continuum modelling, particularly with \nustar\ data where much of the signal is at high energies.

We confirm the difference in $\Gamma$ between the reflection and continuum spectra found by \citet{Fuerst15}, finding $\Gamma\sim2.1$ for the reflection and $\sim2.4$ for the continuum (calculated by replacing the Comptonized continuum with a cut-off power-law, and ignoring the data below 3~keV where $comptt$ turns over and cannot be replicated with a power-law). These measurements are inconsistent at $>3\sigma$, and the difference is similar in size to that found by \citet{Fuerst15}. We are unable to find a good fit when $\Gamma$ is fixed between the two spectral components. There are several potential reasons for this, but perhaps the most likely is systematic effects in the reflection modelling. There are several secondary effects that could affect the broad-band shape of the reflection spectrum, altering $\Gamma$, the iron abundance, and cut-off energy. The assumed density of the latest generation of reflection models is $10^{15}$~cm$^{-3}$, which is appropriate for AGN but should be orders of magnitude larger in XRBs. Standard reflection models do not include a variable low energy cut-off, and so overpredict the flux at low energies, which may cause the slope to become harder to compensate. A further point to consider is the impact of the thermal emission from the disk on the total reflection spectrum. There have been models which include this in the past \citep{Ross07}, but these include the contribution from the black body in the reflection spectrum, so are unsuitable for use with \textsc{kerrbb}. We cannot say definitively whether the difference in $\Gamma$ is real or due to systematic effects, but it should be possible to determine with more sophisticated modelling. We do not expect any of these effects to strongly impact the shape of the line profile, so our results on the spin and inclination should be robust.
If the difference in $\Gamma$ is real, it is likely due to the geometry of the corona. In particular, outflowing coronal models predict that different $\Gamma$ values would be seen by the disk and observers at infinity \citep[e.g.][]{Beloborodov99}.

\section{Conclusions}
\label{section_conclusions}

We present observations of the transient X-ray binary GX~339-4 in the very high state, using \nustar\ and \swift . With the high sensitivity and lack of pile-up available with \nustar , combined with the broad-band spectrum made possible by the addition of the \swift\ XRT, we are able to use both reflection and continuum fitting methods to find measurements of several key parameters. All errors below are statistical only.
\begin{itemize}
\item Using reflection fitting, we find a high spin ($a=0.95^{+0.02}_{-0.08}$) and low inclination ($i=30$\textdegree$\pm1$), which agree well with previous measurements using reflection. 
\item By combining the reflection and continuum fitting methods we are able to estimate the mass and distance of GX~339-4 using the X-ray spectrum alone, finding a mass of $9.0^{+1.6}_{-1.2}M_\odot$ and distance of $8.4\pm0.9$~kpc.
\item We explore the effect of mass and distance on the estimates of $a$ and $i$ from continuum fitting, finding a band of solutions consistent with the results from reflection.
\item We confirm the result of \citet{Fuerst15} that a different value of $\Gamma$ between the reflection and continuum spectra is required, at a high level of significance.
\end{itemize}

\section*{Acknowledgements}
We are grateful to the referee for detailed and thoughtful comments that have significantly improved the paper.
MLP acknowledges financial support from the STFC. PR acknowledges financial contribution from ASI-INAF I/004/11/0 and ASI-INAF I/037/12/0.
This work made use of data from the \nustar\ mission, a project led by the California Institute of Technology, managed by the Jet Propulsion Laboratory, and funded by NASA. This research has made use of the \nustar\ Data Analysis Software (NuSTARDAS) jointly developed by the ASI Science Data Center and the California Institute of Technology. 

\bibliographystyle{mn2e}

\end{document}